**RESEARCH ARTICLE**
10.1002/2015JA022050

# Seasonal and diurnal variations in AMPERE observations of the Birkeland currents compared to modeled results

**J. C. Coxon[1,2], S. E. Milan[1,3], J. A. Carter[1], L. B. N. Clausen[4], B. J. Anderson[5], and H. Korth[5]**

[1]Department of Physics and Astronomy, University of Leicester, Leicester, UK, [2]Department of Physics and Astronomy, University of Southampton, Southampton, UK, [3]Birkeland Centre for Space Science, University of Bergen, Bergen, Norway, [4]Department of Physics, University of Oslo, Oslo, Norway, [5]The Johns Hopkins University Applied Physics Laboratory, Laurel, Maryland, USA

**Abstract** We reduce measurements made by the Active Magnetosphere and Planetary Electrodynamics Response Experiment (AMPERE) to give the total Birkeland (field-aligned) current flowing in both hemispheres in monthly and hourly bins. We analyze these totals using 6 years of data (2010–2015) to examine solar zenith angle-driven variations in the total Birkeland current flowing in both hemispheres, simultaneously, for the first time. A diurnal variation is identified in the total Birkeland current flowing, consistent with variations in the solar zenith angle. A seasonal variation is also identified, with more current flowing in the Northern (Southern) Hemisphere during Bartels rotations in northern (southern) summer. For months close to equinox, more current is found to flow in the Northern Hemisphere, contrary to our expectations. We also conduct the first test of the Milan (2013) model for estimating Birkeland current magnitudes, with modifications made to account for solar contributions to ionospheric conductance based on the observed variation of the Birkeland currents with season and time of day. The modified model, using the value of $\Phi_D$ averaged by Bartels rotation (scaled by 1.7), is found to agree with the observed AMPERE currents, with a correlation of 0.87 in the Northern Hemisphere and 0.86 in the Southern Hemisphere. The improvement over the correlation with dayside reconnection rate is demonstrated to be a significant improvement to the model. The correlation of the residuals is found to be consistent with more current flowing in the Northern Hemisphere. This new observation of systematically larger current flowing in the Northern Hemisphere is discussed in the context of previous results which suggest that the Northern Hemisphere may react more strongly to dayside reconnection than the Southern Hemisphere.

## 1. Introduction

Field-aligned currents electrodynamically connect the ionosphere to the magnetopause and the inner magnetosphere, transmitting stresses through the magnetosphere (shown in Figure 1). These field-aligned currents are also known as Birkeland currents, since they were first proposed by *Birkeland* [1908, 1913]. Sixty years after they were proposed, the first measurements of these currents were made [*Zmuda et al.*, 1966; *Cummings and Dessler*, 1967]. A decade later, the Birkeland currents were found to comprise regions 1 and 2 Birkeland currents above the auroral ionosphere, with region 1 (R1) currents lying poleward of the region 2 (R2) currents and closing through horizontal Pedersen currents in the ionosphere *Iijima and Potemra* [1976a, 1976b, 1978]. R1 currents were shown to map to the magnetopause and the magnetotail, and observed to flow upward (downward) in the dusk (dawn) sector. R2 currents flow in the opposite sense in the dawn and dusk sectors and were shown to map to the partial ring current in the inner magnetosphere [e.g., *Cowley*, 2000, and references therein]. It was shown by *Fujii et al.* [1981] that the currents show a seasonal dependence, which is connected to seasonal variations in ionospheric conductance, and more recently, *Coxon et al.* [2014a, 2014b] showed that the currents are consistent with driving by magnetic reconnection on the dayside of Earth.

There are two drivers of ionization in the *E* region of the ionosphere and therefore two drivers of conductance: auroral particle precipitation and solar extreme ultraviolet radiation [*Robinson and Vondrak*, 1984]. As a result of the latter, the conductance of the ionosphere is seasonal and therefore asymmetric between the two hemispheres, peaking in the summer when the EUV contribution is at its highest. Additionally, there is a diurnal effect caused by the displacement of the geomagnetic pole from the rotational pole of the planet, such that the EUV contribution will be at its largest at local noon (which is at 16:49 UT (04:49 UT) for the northern (southern) geomagnetic pole in the current epoch).

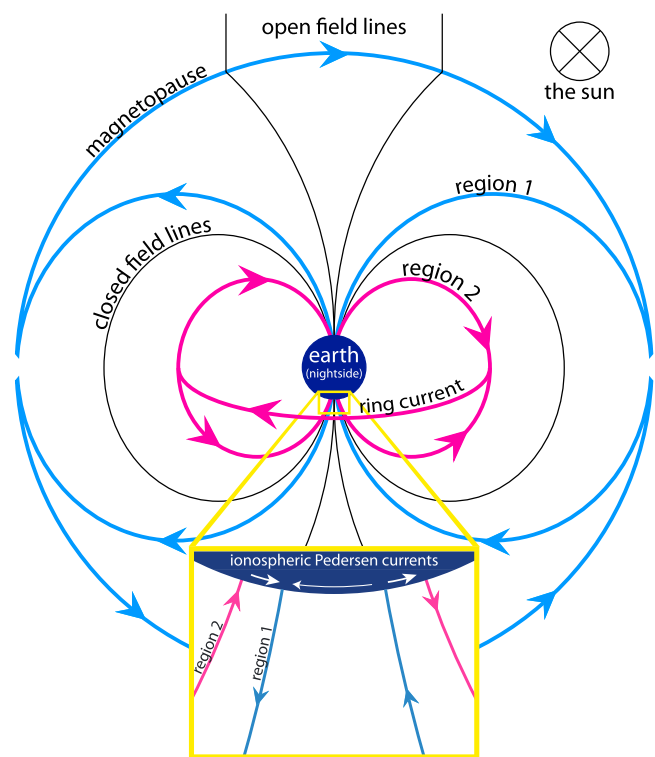

**Figure 1.** A diagram drawn as if Earth was eclipsing the sun. It shows the region 1, region 2, magnetopause, and ring currents as well as illustrates the location of open and closed terrestrial magnetic field lines. It can be seen from this image how the region 1 current sheet corresponds to the open/closed field line boundary, or OCB.

Previously, attempts have been made to find the relationship between the solar zenith angle $\chi$ and the conductance $\Sigma$ [*Robinson and Vondrak*, 1984; *Moen and Brekke*, 1993, and references therein] by using radar measurements; attempts have also been made to explore this connection by observation of the Birkeland currents directly [*Fujii and Iijima*, 1987]. Previous authors have also noted that there are hemispherical asymmetries in the Birkeland currents which arise from this variation in ionospheric conductance [*Fujii et al.*, 1981; *Ohtani et al.*, 2005; *Wang*, 2005]. Observations of Birkeland currents have been corroborated by MHD models which also suggest a conductance-driven asymmetry in Birkeland current [*Wiltberger et al.*, 2009], but observations have also been made which suggest that this effect may not be entirely attributable to conductance [*Chisham et al.*, 2009]. Recently, such estimates of ionospheric conductance have been used in conjunction with field-aligned current densities and radar data in order to attempt to quantify high-latitude ionospheric electrodynamics [*Cousins et al.*, 2015a, 2015b].

In order to fully understand the Birkeland currents, it is necessary to consider the Dungey cycle, in which interplanetary field lines interconnect with terrestrial field lines in order to create open magnetic flux, which convects tailward across the polar cap and is subsequently closed on the nightside before circulating back to the dayside to complete the cycle [*Dungey*, 1961]. Originally envisioned as a static phenomenon, *Cowley and Lockwood* [1992] proposed that this process was in fact dynamic and that observations of the changing size of the polar cap could be attributed to differences between dayside and nightside reconnection rates. This has become known as the expanding/contracting polar cap (ECPC) paradigm and has been corroborated by numerous observations [e.g., *Milan et al.*, 2007, 2012].

*Milan* [2013, hereafter M13] constructed an electrodynamic model which estimated the Birkeland currents flowing by extending models of the convection driven by the ECPC paradigm [*Freeman and Southwood*, 1988]. In the M13 model, the magnetic reconnection rates $\Phi_D$ and $\Phi_N$ are specified and then the model current flows are calculated (given assumed Hall and Pedersen conductances). The currents are found to depend on the transpolar voltage (the average of the dayside and nightside reconnection voltages), with stronger currents found on the dayside when dayside reconnection dominates and stronger currents found on the nightside when nightside reconnection dominates. It has been demonstrated that Birkeland current magnitudes are consistent with driving by dayside and nightside reconnection [*Coxon et al.*, 2014a]. During substorms, more current flows [*Clausen et al.*, 2013a, 2013b; *Coxon et al.*, 2014b; *Sergeev et al.*, 2014a, 2014b], although *Murphy et al.* [2012] found a decrease before auroral substorm onset, and some authors have noted that the spatial structure of substorms is more complex than previously thought [*Murphy et al.*, 2013; *Forsyth et al.*, 2014]. It has also been found that the Birkeland current ovals move equatorward and poleward during dayside and nightside reconnection, consistent with the ECPC paradigm [*Clausen et al.*, 2012; *Coxon et al.*, 2014b; *Carter et al.*, 2016].

The work in this paper builds on the previous examinations of the dependence of ionospheric conductance and current flow on solar zenith angle. In this paper, the effect of ionospheric conductance is examined using AMPERE (the Active Magnetosphere and Planetary Electrodynamics Response Experiment, outlined in section 2.1), which allows for the current flowing in both hemispheres over 6 years to be analyzed in the context of diurnal and seasonal variations, an enhancement upon previous studies. The work presented in this paper also allows for the dependence of field-aligned currents on solar zenith angle to be investigated simultaneously in both hemispheres, something which has not previously been possible. Additionally, we extend the M13 model, including the ionospheric conductance $\Sigma$ based on the calculations of *Moen and Brekke* [1993], to incorporate the effects of solar flux $F_{10.7}$ and the solar zenith angle. This allows for the current magnitudes to be modeled as a function of solar flux, solar zenith angle, and dayside reconnection rate for the first time. The correlation between the magnitudes $J$ (inferred from AMPERE) and dayside reconnection rate $\Phi_D$ is calculated. We also calculate the correlation between $J$ and the result of the M13 model, before discussing the improvements we gain by enhancing the model with the solar flux and solar zenith angle.

## 2. Sources of Data Utilized

### 2.1. AMPERE and Derived Products

Engineering magnetometer data from the Iridium® telecommunications satellite network are used to construct global maps of the Birkeland current systems as part of the AMPERE experiment [*Anderson et al.*, 2014]. This constellation of satellites comprises 11 spacecraft in six orbital planes for a total of 66 satellites in polar orbits of 780 km altitude which take 104 min to complete. Measurements are provided along 12 meridians of magnetic local time (two conjugate hours of MLT in each orbital plane). The magnetic perturbations measured by the Iridium® network were used by *Anderson et al.* [2000] to deduce field-aligned current magnitude, and more recently, AMPERE has improved on this method to give large-scale Birkeland currents at a cadence of 10 min. In this study, data from January 2010 to December 2015 are used.

In order to obtain the total current flowing in either hemisphere, the current signatures along a given meridian of MLT are considered. Any current value lower than 0.2 μA m$^{-2}$ is neglected following the findings of *Clausen et al.* [2012], and any current signature further than 30° from the geomagnetic pole is also ignored. The total upward current is determined by integrating under the positive current signatures and neglecting the downward current signatures (and vice versa for the total downward current). The total current flowing in that hemisphere is the sum of the upward and downward currents across all MLT $J = |J_\uparrow| + |J_\downarrow|$.

### 2.2. OMNI and Derived Products

The OMNI data set provides time series of solar wind parameters propagated to their impact on the bow shock [e.g., *King*, 1991; *Papitashvili et al.*, 2000, and references therein]. Data from OMNI are used to estimate the dayside reconnection rate $\Phi_D$, using the expression by *Milan et al.* [2012]:

$$\Phi_D = L_{\text{eff}}(V_X) V_X B_{YZ} \sin^{\frac{9}{2}} \left( \frac{\theta}{2} \right). \tag{1}$$

In the above equation $L_{\text{eff}}(V_X)$ is an effective length scale, given by

$$L_{\text{eff}}(V_X) = 3.8 R_E \left( \frac{V_X}{4 \times 10^5 \text{ms}^{-1}} \right)^{\frac{1}{3}}, \tag{2}$$

and $B_{YZ}$ is the transverse component of the interplanetary magnetic field (IMF), given by

$$B_{YZ}^2 = B_Y^2 + B_Z^2. \tag{3}$$

$V_X$ is the solar wind speed, $\theta$ is the clock angle between the IMF vector projected into the GSM $Y$-$Z$ plane and $Z$ axis, and $R_E$ is the radius of Earth.

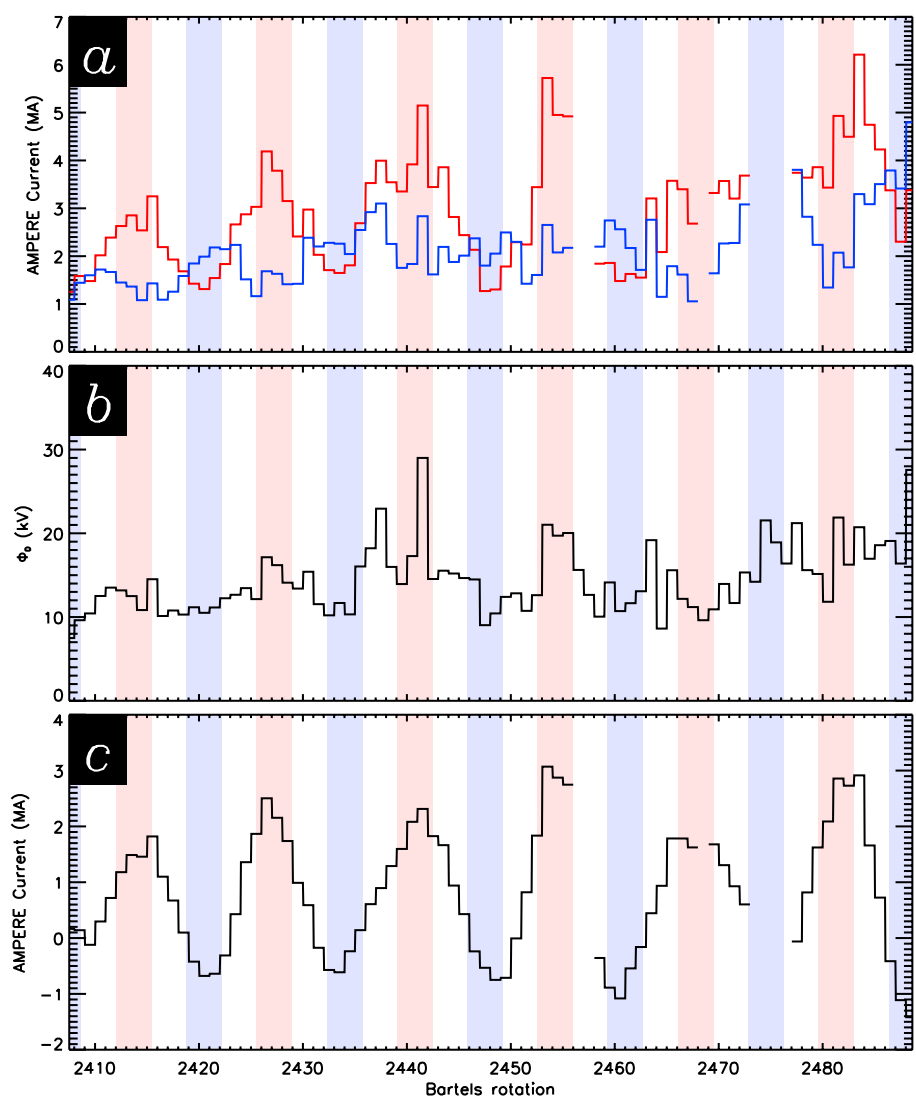

**Figure 2.** A diagram showing Bartels rotation on the x axis. Plotted on the y axis, from top to bottom: (a) the current magnitudes $J_N$ (in red) and $J_S$ (in blue); (b) the dayside reconnection rate $\Phi_D$; (c) the difference in current magnitude $J_N - J_S$. Each variable is plotted as the mean per Bartels rotation. A light red background indicates May–July, and a light blue background indicates November–January. The current magnitudes plotted are the sum of the total downward and total upward current inferred from AMPERE.

## 3. Birkeland Current Strengths 2010–2015

Birkeland currents and their strengths are measured over a period of 6 years using AMPERE data reduced using the techniques described in section 2.1. The total Birkeland current flowing is recorded in the Northern Hemisphere $J_N$ and in the Southern Hemisphere $J_S$. *Coxon et al.* [2014a, 2014b] have shown that magnetic reconnection and substorms are very important factors in driving Birkeland current strengths and determining their size and position, but this study aims to separate the effect of driving by reconnection from the effect of ionospheric conductance in order to examine the effect of seasonal and diurnal variations on the system.

### 3.1. Mean Birkeland Current Strengths Per Bartels Rotation

Figure 2a shows the mean per Bartels rotation of the Birkeland current strengths measured by AMPERE plotted against the number of days since 1 January 2010 in the Northern and Southern Hemispheres, plotted in red and blue, respectively). Points where there is a gap in either line are due to a lack of availability of AMPERE data during that Bartels rotation. Diurnal and seasonal variations in solar illumination controlled by the solar zenith angle will give rise to variations in ionospheric conductance.

Figure 2b shows the mean per Bartels rotation of the dayside reconnection rate $\Phi_D$, for the period 2010–2015. The mean $\Phi_D$ varies significantly during this time. Peaks in the dayside reconnection rate are observed in or near to the summers in the Northern Hemisphere (indicated by light red stripes) in 2010–2013, with troughs in the dayside reconnection rate observed in the winters in the Northern Hemisphere (indicated by light blue stripes) in 2010–2012. However, the dayside reconnection rate is less well ordered by season after the summer of 2013, with peaks and troughs no longer seen coincidental with season.

The dayside reconnection rate $\Phi_D$, which controls the ionospheric convection, has a semiannual variation [*Russell and McPherron*, 1973] which leads to more activity at the equinoxes; other than this, $\Phi_D$ is not expected to have a seasonal or diurnal variation. Thus, it is proposed that there will be a seasonal and diurnal variation of the Birkeland currents which is solely due to the solar illumination effect on the ionospheric conductance. The presence of such an effect can be verified by plotting $J_N - J_S$ in Figure 2c. $J_N$ is up to 3 MA higher than $J_S$ during summer in the Northern Hemisphere, whereas $J_S$ is approximately 1 MA higher than $J_N$ during summer in the Southern Hemisphere.

In order to identify the seasonal variation in the Birkeland currents, we consider the change in the conductance with season. Instead of using the local conductance $\Sigma$ [*Robinson and Vondrak*, 1984; *Fujii and Iijima*, 1987], we assume that the total amount of Birkeland current flowing can be described by the dayside reconnection rate (a voltage) multiplied by some term $\Xi$. We make this assumption because Birkeland current strength has been previously shown to be well correlated with $\Phi_D$ [*Coxon et al.*, 2014a]. $\Xi$ is so called to differentiate it from $\Sigma$ and will vary with season and will have units of conductance; this assumes that the seasonal effect on the Birkeland currents can be quantified by a single number. This assumption yields the following relation describing total Birkeland current:

$$J = \Phi_D \Xi. \tag{4}$$

If the data are averaged over a long enough period of time, we can employ $\Phi_D$ in isolation without explicitly considering $\Phi_N$, because on the timescale of a Bartels rotation, $\Phi_D = \Phi_N$ [*Cowley and Lockwood*, 1992]. Equation (4) refers to the total Birkeland current flowing, which can then be subdivided into Northern and Southern Hemisphere current flow $J_N$ and $J_S$, respectively.

We assume that the conductance $\Xi$ can be quantified as a sinusoid that has a period of 1 year, equal to $y_0$ on the 79th day of 2010, which was the vernal equinox in that year. Thus,

$$t = \frac{2\pi(d - 79)}{365.25} \tag{5}$$

$$\Xi(t) = y_0 \pm y_a \sin(t) \tag{6}$$

can be written where $y_0$ is the background conductance, $y_a$ is the variation in conductance due to seasonal effects (the amplitude of the sinusoid), and $d$ is the number of days since 1 January 2010. In the Southern Hemisphere, $y_a$ is expected to be negative, as opposed to the positive amplitude expected in the Northern Hemisphere.

Equation (4) implies that $\Phi_D = J/\Xi(t)$. Consequently, we can compare the estimate of $\Phi_D$ from OMNI data with an estimated $\Phi_D^*$ calculated by dividing current magnitude $J$ by the sinusoid $\Xi(t)$, and this means that the form of $\Xi(t)$ that provides the best correlation between the two can be determined using equation (6). By exploring values of $y_0$ and $y_a$, we can find the form of equation (6) that gives the best fit between $\Phi_D$ and $J/\Xi(t)$ for each hemisphere to find the conductance in the north $\Xi_N$ and the south $\Xi_S$. To this end, we obtained the values of $y_a$ and $y_0$ that give the best correlation coefficient using a brute force approach, giving the results

$$\Xi_N(t) = 202.6 + 54.0\sin(t) \text{ mho}, \tag{7}$$

$$\Xi_S(t) = 154.4 - 50.6\sin(t) \text{ mho}. \tag{8}$$

The fact that $y_0$ in the Northern Hemisphere is 48.2 mho higher than in the Southern Hemisphere indicates that the Northern Hemisphere experiences consistently larger Birkeland current magnitudes than the Southern Hemisphere across the 6 years plotted. This is consistent with the observation that the southern magnitudes are 1 MA larger than the northern magnitudes during the southern summer, when in the northern summer the northern magnitudes are 3 MA larger than the southern magnitudes.

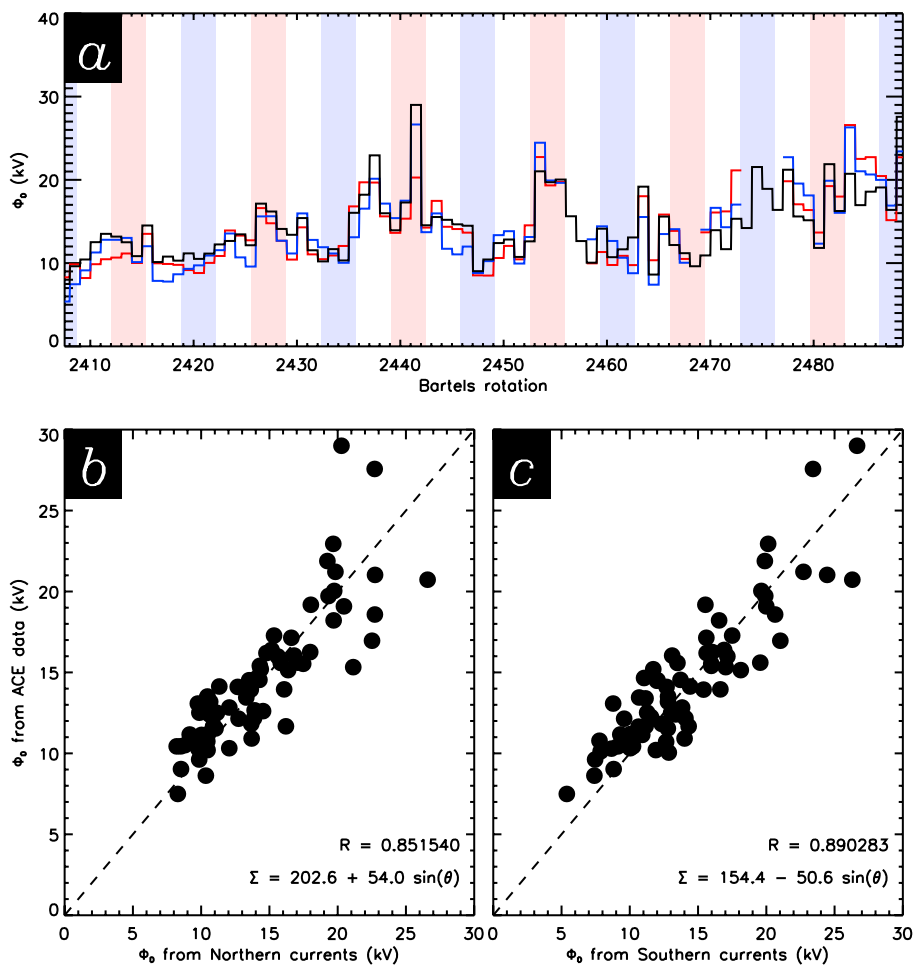

**Figure 3.** (a) A diagram showing the $\Phi_D$ plot shown in Figure 2b (Bartels rotation versus $\Phi_D$) with the estimated $\Phi_D$ based on $J_N$ (red) and $J_S$ (blue) overplotted. (b) A plot of $\Phi_D$ showing the AMPERE-based prediction against the OMNI measurement for the Northern Hemisphere. (c) As in Figure 3b for the Southern Hemisphere.

Figure 3 shows the result of $\Phi_D^*$. Figure 3a shows the mean values of $\Phi_D$ previously plotted in Figure 2b with the northern and southern estimates of $\Phi_D$ (using the solutions presented in equations (7) and (8)) overlaid in red and blue, respectively. Figure 3 also shows the measured $\Phi_D$ plotted against the estimated $\Phi_D$ in the Northern (Figure 3b) and Southern (Figure 3c) Hemispheres. In both hemispheres, a Pearson correlation coefficient of at least 0.85 is calculated, indicating a strong correlation.

### 3.2. Mean Diurnal Birkeland Current Strengths

We now turn to examine diurnal variations expected due to solar zenith variations, and Figure 4 shows mean Birkeland current magnitudes $J_N$ (Figure 4a) and $J_S$ (Figure 4b), plotted as colors on a graph of Bartels rotation (on the $x$ axis) against the hour of UT (on the $y$ axis). The dayside reconnection rate $\Phi_D$ is plotted with the same format in Figure 4c. In order to make the diurnal variation more clear, we also plot the hourly mean variation in each of these three variables over the 6 year period to the right of these plots (Figures 4d–4f).

In order to split the data into the 24 h of the day, we bin the Birkeland current magnitudes by taking the intervals which have start times in a given hour (such that the 00–01 UT bin contains results from the range $00:00 \le UT < 01:00$). We then use each of these 24 bins to take the mean average across every day in a given Bartels rotation to find the magnitude of the Birkeland currents for a given combination.

The seasonal variation as described in section 3.1 is also observed in Figure 4, seen in the columns. However, this method of presenting the data also shows a diurnal variation, which is highlighted by the hourly mean averages shown in Figures 4d–4f. In the Northern Hemisphere, the maximum current strength is observed at 16–17 UT, whereas the minimum current strength is observed at 00-01 UT. In the Southern Hemisphere, the maximum current strength is observed at 05–06 UT, whereas the minimum is at 16–17 UT. There is not an apparent diurnal variation in $\Phi_D$, indicating that any observed diurnal variation cannot be explained by driving dayside reconnection.

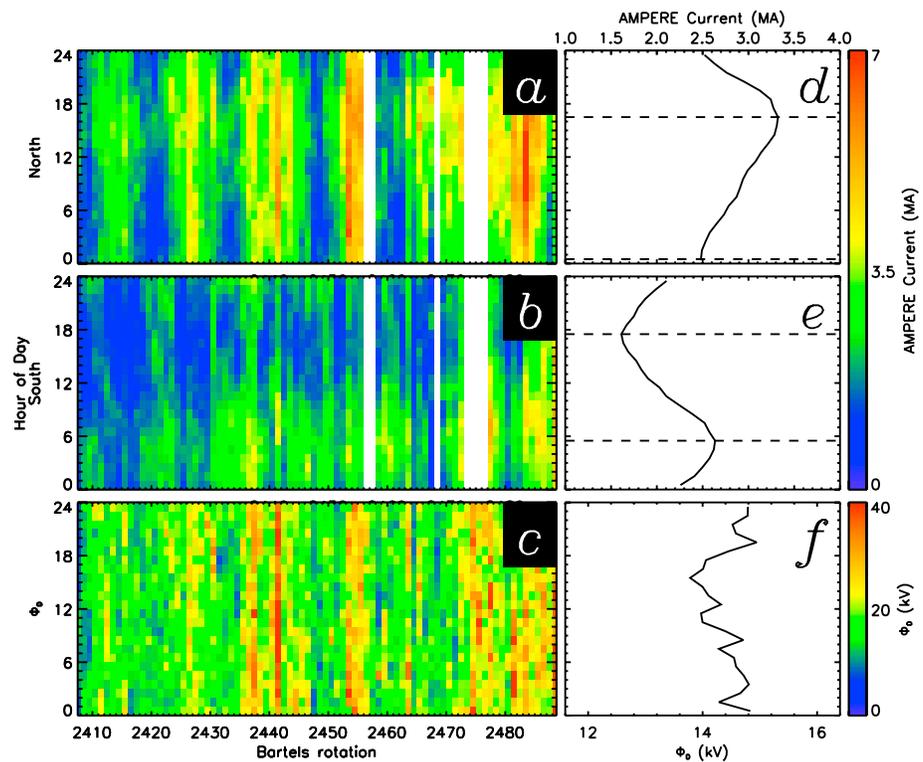

**Figure 4.** (left column) Diagrams plotted with Bartels rotation on the x axis and hour of day on the y axis. On the z axis, from top to bottom: (a) the northern current magnitude $J_N$; (b) the southern current magnitude $J_S$; (c) the dayside reconnection rate $\Phi_D$. (right column) Diagrams with hour of day on the y axis and, on the x axis, (d) $J_N$, (e) $J_S$, and (f) $\Phi_D$.

## 4. Comparison With the Model of *Milan* [2013]

The results in section 3 indicate that the magnitude of the field-aligned currents is controlled by the magnetic flux transport in the Dungey cycle, quantified by the dayside reconnection rate $\Phi_D$, combined with the ionospheric conductance partly controlled by solar illumination. Seasonal and diurnal variations in solar illumination of the polar regions result in similar variations in the field-aligned current magnitudes. To verify this conclusion, we employ a simple model of the polar electrodynamics to predict the expected current magnitudes for given levels of solar illumination and magnetic flux transport driven by $\Phi_D$.

We used the M13 model, based upon the ECPC paradigm, to calculate an electrostatic potential pattern (associated with the ionospheric convection pattern) consistent with given rates of dayside and nightside reconnection and hence expansion or contraction rate of the polar cap. It was assumed that the polar cap had a low level of ionospheric conductance, and the auroral zone, coincident with the region of convection return flow, formed a ring of higher conductance surrounding this. The electric field model and conductance pattern were combined to determine the Pedersen current flow in the ionosphere; downward (upward) field-aligned currents are associated with regions of divergence (convergence) in the Pedersen current flow. That model used a particularly simple conductance model as it allowed analytical expressions for the FAC magnitudes to be derived; in the present modeling, we modify the M13 model to include the solar contribution to conductance.

Our aim is to model the observed Bartels-averaged field-aligned current magnitudes using the Bartels-averaged dayside reconnection rate $\Phi_D$ as the input, for comparison to Figure 3. We assume in section 3.1 that $\Phi_D = \Phi_N$; hence, the magnetic flux transport rate quantified by the cross-polar cap potential $\Phi_{PC}$ is equal to $\Phi_D$. We also assume that the polar cap is circular and centered on the geomagnetic pole, such that the polar cap boundary is located at a magnetic colatitude of 15°, and that the auroral zone and return flow regions are collocated and 7° of latitude in width. Using the M13 model, the electrostatic potential pattern is determined alongside associated poleward and azimuthal components of the electric field for a given Bartels-averaged value of $\Phi_{PC}$, evaluated on a grid with a latitudinal spacing of 1° and a longitudinal spacing of 2°. It should be noted that the $\Phi_{PC}$ values have been scaled by 1.7, which we will discuss in more detail in section 5.3.

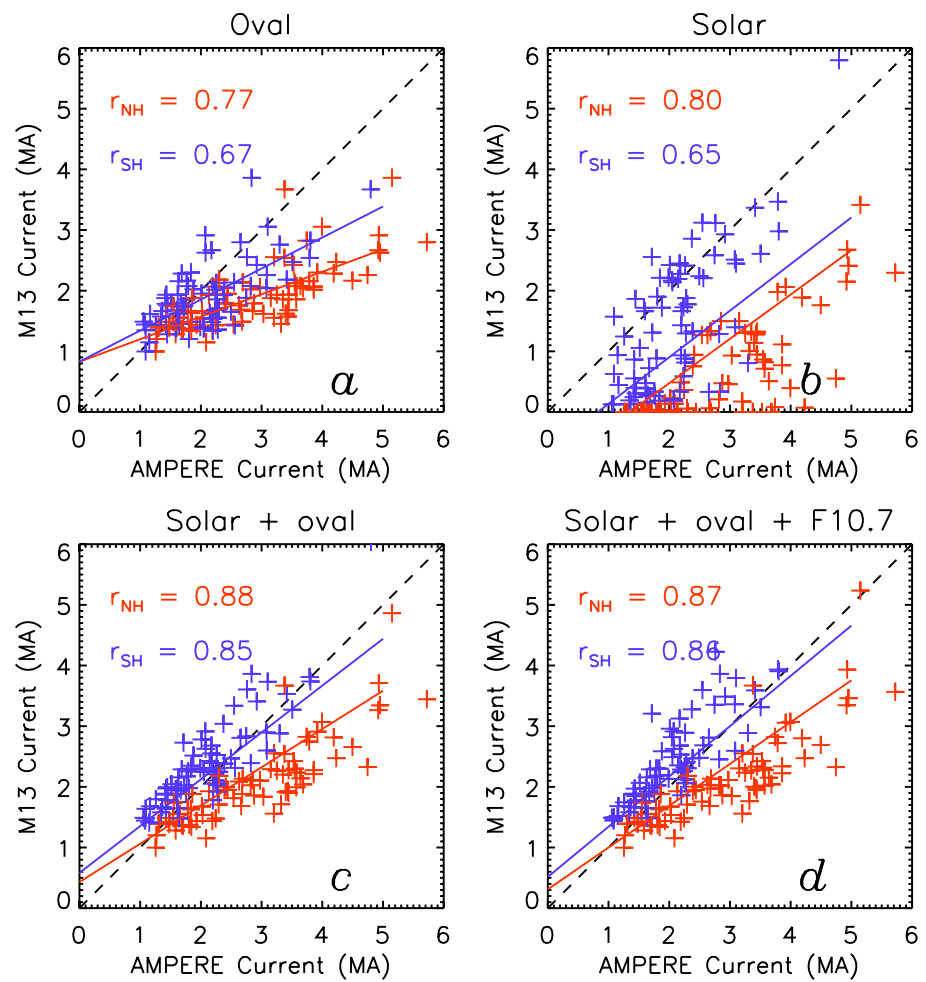

**Figure 5.** Plots showing observed field-aligned current strength (*x*) versus modeled field-aligned current strength (*y*), in MA. Red corresponds to the Northern Hemisphere; blue, to the Southern Hemisphere. The dashed black line in each panel is the line of unity, and the solid colored line is the linear least squares fit between the observed and modeled currents. In each panel, the modeled current magnitudes are calculated differently: (a) calculated solely using the average conductance in the auroral oval; (b) calculated solely using the solar contribution to conductance; (c) combining the auroral oval conductance with the solar input to conductance; (d) combining the two and taking account of $F_{10.7}$.

For simplicity, we assume that the ionospheric conductance is uniform within the auroral zone region, using values of $\Sigma_P = 7$ mho and $\Sigma_H = 12$ mho, taken as typical values from the $K_p = 3$ conductance model of *Hardy et al.* [1987]. We evaluate the solar contribution to conductance using the expressions of *Robinson and Vondrak* [1984] and *Moen and Brekke* [1993], which both depend on $F_{10.7}$ flux and solar zenith angle; we find that the two models give near-identical results, and so we use the model of *Moen and Brekke* [1993]; where we have not included the $F_{10.7}$ parameter we simply set $F_{10.7} = 100$ solar flux units. We calculate the poleward and azimuthal components of both the Hall and Pedersen currents in each grid cell of the modified model. The divergence of these currents between adjacent grid cells then yields the spatial distributions of upward and downward FACs. We sum the upward and downward currents in the same manner as the observations, allowing for the observations to be compared to the modeled results in Figure 5.

Figure 5 has four panels which show different contributions to the modeled currents from the solar-produced conductances, with and without accounting for changes in $F_{10.7}$ flux and auroral oval conductance. The solid, colored lines show linear least squares fits between the observed and modeled currents.

Figure 5a shows the currents modeled only with a constant auroral oval conductance: the modeled currents have a correlation with the observed currents in the Northern Hemisphere $r_{NH} = 0.77 \pm 0.05$, but in the Southern Hemisphere $r_{SH} = 0.67 \pm 0.07$. Figure 5b shows the currents modeled only with solar contributions to conductance included; in this case, $r_{NH} = 0.80 \pm 0.04$ and $r_{SH} = 0.65 \pm 0.10$. Figure 5c shows the currents modeled using both auroral and solar contributions to conductance in Figures 5a and 5b; the correlations increase to $r_{NH} = 0.88 \pm 0.02$ and $r_{SH} = 0.85 \pm 0.04$. Finally, Figure 5d shows the same contributions as in Figure 5c but with $F_{10.7}$ set to the mean average value per Bartels rotation. The correlations in this final case are $r_{NH} = 0.87 \pm 0.03$ and $r_{SH} = 0.86 \pm 0.04$. The errors quoted for each correlation coefficient were estimated by bootstrapping 1000 copies of the AMPERE and M13 magnitudes and taking the standard deviation of the result. Incorporating both auroral and solar contributions increases the correlation coefficients seen in both hemispheres; this result will be analyzed in more detail later.

# 5. Discussion

Previous work has shown that Birkeland current magnitude $J$ is driven by dayside reconnection rate $\Phi_D$ [*Coxon et al.*, 2014a; *Anderson et al.*, 2014] and is also related to substorm processes in the magnetotail [*Coxon et al.*, 2014b]. Previous studies have linked conductance to Birkeland current strengths [*Fujii and Iijima*, 1987] and examined the conductances of other current systems [*Moen and Brekke*, 1993]. In the discussion below, we quantify the significance of the inclusion of a conductance term to the relationship between $J$ and $\Phi_D$ to illustrate the improvement yielded by the method described in this paper. We also discuss a potential bias in the AMPERE data set toward generally larger Northern Hemisphere values.

## 5.1. Seasonal Variation in Conductance

Figure 2 shows a clear seasonal variation in the current magnitude in the Northern Hemisphere $J_N$ and in the Southern Hemisphere $J_S$. This seasonal variation is a total of 4.2 MA, with $J_N - J_S = 3.1$ MA at the height of the northern summer and $J_N - J_S = -1.1$ MA during the southern equivalent. This apparently suggests that the Northern Hemisphere enjoys proportionally larger current magnitudes as a result of season, compared to the Southern Hemisphere. The observed larger northern currents might be a coincidental bias over the period of data shown but would also be consistent with a bias toward larger Northern Hemisphere values in the AMPERE data set. This would be in agreement with some of the conclusions drawn in section 3.1. However, the peaks in the mean $\Phi_D$ are located during northern summers and the troughs are located in southern summers, which might also provide an explanation for why stronger currents are found in the Northern Hemisphere.

Comparing the Pearson correlation coefficient between $J$ and $\Phi_D$ assuming no seasonal variation in conductance (the null hypothesis) allows insight into the variation of $J$ with season. In the Northern Hemisphere, $R = 0.77$, and in the Southern Hemisphere $R = 0.67$. We conclude that the apparent correlation in the Northern Hemisphere is due to the coincidental occurrence of high $\Phi_D$ during the Northern Hemisphere summer. We measure correlation coefficients for the data of $R = 0.85$ and $R = 0.89$ in the Northern and Southern Hemispheres, respectively, yielding improvements $\Delta R = 0.08$ in the north and $\Delta R = 0.22$ in the south. The difference is small in the Northern Hemisphere but more pronounced in the Southern Hemisphere. We conducted a $Z$ test [*Cohen et al.*, 2003] to determine the $p$ value of the increase in correlation and found that the increase is significant at the 95% level in both hemispheres, showing a clear seasonal effect in the data.

## 5.2. Variation in Conductance With $\chi$

In order to examine the effect of the solar zenith angle on this result, calculations of that angle $\chi$ at the geomagnetic pole are considered. An estimate of the position of the geomagnetic pole in 2010 is used, with geographic latitude $\Phi = 80.1°$ in both hemispheres and longitude $\lambda_N = -72.2°$ in the Northern Hemisphere and $\lambda_S = 107.8°$ in the Southern Hemisphere [*Finlay et al.*, 2010]. The value of $\chi$ at the geomagnetic pole across the years 2010–2015 is shown in Figures 6a and 6b, with the diurnal variations plotted in Figures 6c and 6d and the variation with Bartels rotation depicted in Figures 6e and 6f. Figures 6a, 6c and 6e show the solar zenith angle in the Northern Hemisphere $\chi_N$, whereas Figures 6b, 6d and 6f show the Southern Hemisphere $\chi_S$.

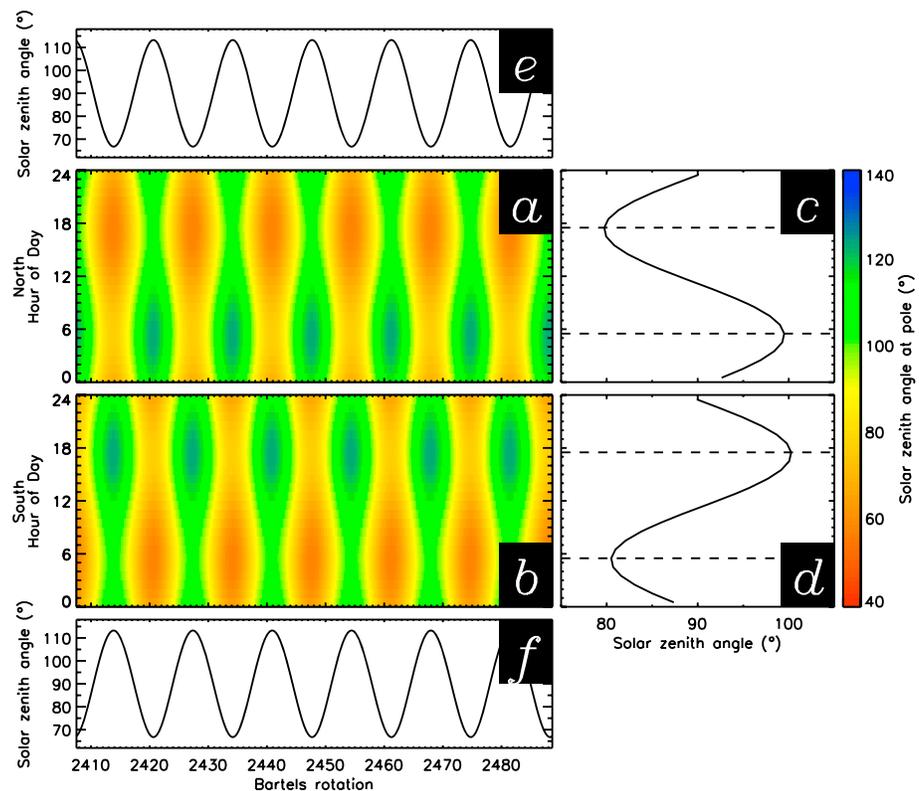

**Figure 6.** (a and b) Diagrams plotted with Bartels rotation on the x axis and hour of day on the y axis. On the z axis, solar zenith angle χ at the geomagnetic pole in the Northern and Southern Hemispheres. (c and d) Plots of χ against hour of day for the Northern and the Southern Hemispheres. (e and f) Plots of χ against Bartels rotation for the Northern and Southern Hemispheres.

Turning points in $\chi$ (Figures 6e and 6f) are seen coincident with turning points in the current magnitudes (Figure 2c). It can be seen by examination of Figure 2c that each turning point occurs during the appropriate season.

In terms of diurnal current magnitude variations, turning points are seen in both the morning and afternoon. As noted in section 3.2, no diurnal variation in $\Phi_D$ is observed; this indicates that any diurnal effect must be a result of other factors, and as such the diurnal variation of $\chi$ is plotted in Figures 6c and 6d, showing turning points at 04–05 UT and 16–17 UT. In the Northern Hemisphere, a turning point in the average current magnitude is observed at 16–17 UT, and in the Southern Hemisphere a turning point is observed at 17–18 UT. These turnings are coincident with the turning point in $\chi$, consistent with the view of diurnal variations in conductance affecting current magnitudes. Examining the morning turning points, the Southern Hemisphere's maximum current strength $J_S$ is observed at 05–06 UT, which is an hour later than the minimum seen in $\chi_S$ at 04–05 UT. The minimum in $J_N$ is observed at 00–01 UT, 4 h earlier than the corresponding maximum in $\chi_N$ at 04–05 UT.

In the Southern Hemisphere, both turning points were observed an hour later than expected, with a difference in current magnitude of ~0.02 MA between the expected and observed times of turning point. In the Northern Hemisphere, the maximum was observed at the expected time but the minimum was observed 4 h later, at a value 0.17 MA lower than the current magnitude at the expected turning point. What is causing this feature is unclear and should be examined in a future study.

### 5.3. Correspondence With the Model of *Milan* [2013]

Figure 5 shows the correspondence between the observed monthly averages of the field-aligned current magnitudes and those calculated using the modified M13 model. Figures 5a and 5b show that the current magnitudes in the Northern Hemisphere agree relatively well with modeled values using just one contribution (auroral or solar) to the ionospheric conductance; the correlation coefficient in such cases is 0.77 and 0.80, respectively, showing strong correlations. This is in contrast to the correlation coefficients found in the Southern Hemisphere, which are somewhat less strong (0.67 and 0.65). We expected this result; the coincidental correspondence between high average values of $\Phi_D$ and the seasonal variation of the Northern Hemisphere currents means that the currents appear to be better correlated than the Southern Hemisphere currents with a single driver. It should be noted, however, that the margins of error inferred from bootstrapping the data overlap between the two correlation coefficients, such that we cannot state with certainty that the Northern Hemisphere is better correlated with dayside reconnection than the Southern Hemisphere.

Figure 5c shows the modeled currents with contributions from both oval and solar-produced conductance with a fixed $F_{10.7}$, and Figure 5d incorporates the changes in $F_{10.7}$. These modeling results indicate that both auroral and solar contributions to conductance need to be accounted for to accurately model the observed current values. The correlations increase to 0.87 and 0.86 in the Northern and Southern Hemispheres, respectively, and the errors estimated using bootstrapping show that these correlations are statistically significantly improved when compared to the coefficients obtained for the model with a sole driver. We also employed the $Z$ test to compare the correlation of the combined model with the correlations for both the single-input models, finding that the improvement over both single inputs was significant at the 95% level in both hemispheres; this supports the validity of the modified M13 model. Incorporating the input from $F_{10.7}$ had an insignificant effect on the correlation between the AMPERE currents and the currents modeled using the M13 model, indicating that this variable does not have a large effect on the current magnitudes.

However, there are caveats associated with this result. In order to achieve the correct magnitude of modeled currents, we have scaled $\Phi_{PC}$ by 1.7, as noted in section 4. It should be noted that the M13 model is linear, in so much as the electric field in the ionosphere scales in the same way as $\Phi_{PC}$, as do the currents and, hence, the divergences in the currents. Scaling the ionospheric conductances by the same amount would achieve the same result. From the model, it is difficult to tell which is underestimated: $\Phi_{PC}$, the conductances, or a combination of the two. Employing a spatially uniform conductance in the auroral zone is a gross simplification and could lead to discrepancies between the modeled and observed currents. We also expect that the conductance will increase, and the average location of the oval could move to lower latitudes with increasingly active conditions, though these effects are not included in the model.

We also calculate the correlation of the residuals from Figure 5d with reconnection rate. The correlation of the residuals in the Northern Hemisphere is $0.13 \pm 0.2$, and this result is not significant at the 95% significance level; we cannot conclude that there is a correlation with the residual in this hemisphere. In the Southern Hemisphere, there is a correlation of $-0.25 \pm 0.14$, which is significant at the 95% significance level; this weak correlation can be attributed to overestimation of the currents in the Southern Hemisphere and indicates that the disparity between hemispheres is more pronounced at larger reconnection rates. Since the model assumes that the hemispheres are symmetrical (beyond the asymmetry introduced by the solar zenith input to the function of *Moen and Brekke* [1993]), this shows that the model is failing to reproduce the observed hemispherical asymmetry and indicates that the difference between the two hemispheres is larger at larger reconnection rates.

An asymmetry in total electron content (TEC) was depicted in Figure 3 of *Clausen and Moen* [2015], showing the average TEC and average AMPERE currents over the Northern Hemisphere between 2010 and 2012. There may be an asymmetry in the ionospheric conductance in the same sense as the asymmetry in TEC, and as such some asymmetry in the morphologies of the currents would explain the asymmetry in the magnitudes. It has previously been demonstrated [e.g., *Weimer*, 2001; *Carter et al.* 2016, and references therein] that $B_Y$ rotates the current systems; this rotation is in opposite senses between the two hemispheres. As a result, if a systematic bias in $B_Y$ exists in 2010–2015, then the observed asymmetry could also be caused by the asymmetry in TEC. Investigation shows a very weak bias in $B_Y$ during 2010–2015, with a median of $-0.14$ nT; further work is needed to analyze whether this effect is large enough to reproduce the observed effect.

There are other studies which have previously observed asymmetries between the two hemispheres. *Tulunay and Grebowsky* [1987] found that electron densities in the Northern Hemisphere were generally larger than in the Southern Hemisphere, implying a higher conductivity. *Förster et al.* [2007] used Cluster EDI data from *Haaland et al.* [2007] and found that ionospheric convection velocities were on average higher in the Northern Hemisphere than in the Southern Hemisphere, consistent with larger current flow. More recently,

*Förster and Haaland* [2015] found that the cross-polar cap potential $\Phi_{PC}$ was on average 5–7% larger in the Southern Hemisphere. They also found that the ionospheric convection velocity was, on average, larger in the Northern Hemisphere than in the Southern Hemisphere, which would be consistent with higher current flow. *Cnossen and Förster* [2016] calculated daily average ion drift velocities for 2001–2013, 2002–2003, and 2005–2007 (all time, active years and quiet years), finding that the northern averages were always higher than the southern averages during 2001–2013 and 2005–2007 and approximately two thirds of the time in 2002–2003. The high-latitude neutral wind vortices at dawn and dusk also tended to be stronger in the Northern Hemisphere. These hemispherical asymmetries were attributed by the authors to asymmetries in the Earth's magnetic field.

## 6. Conclusions

We have estimated current magnitudes using the relation $J = \Phi_D \Xi$, demonstrating that Birkeland currents measured by AMPERE are seasonally dependent, observing seasonal variations in both hemispheres in the 6 years under discussion. These seasonal variations take the form $y_0 + y_a \sin(t)$, where $y_0$ is the background conductance and $y_a$ is the seasonal variation in conductance; we have quantified them as a variation in conductance of $0.27y_0$ in the Northern Hemisphere and $0.33y_0$ in the Southern Hemisphere, with correlation coefficients $R \geq 0.85$ in both hemispheres. Furthermore, we have shown, over the 6 years, that $J$ also displays diurnal variations consistent with a dependence on solar zenith angle $\chi$.

We have conducted a test of the *Milan* [2013] model, modifying the model to account for variations in solar contribution to ionospheric conductance. This has the benefit of both validating the model and also examining whether the Birkeland currents can be well described by a combination of ionospheric conductance and driving by magnetic reconnection. We note that the cross-polar cap potential was scaled by 1.7, implying that the cross-polar cap potential or the conductances were underestimated. We compared the modeled current magnitudes to the actual values, finding a good correspondence with correlation coefficients of 0.87 and 0.86 (north and south, respectively). We therefore conclude that this model provides a statistically significant improvement over the correlation of current magnitude with either driver in isolation and that this model provides a reasonable estimate of the current magnitudes.

We observe an asymmetry in the current magnitudes in the two hemispheres which is not reproduced in the model, and we note that this could be an effect of asymmetry in the Earth's magnetic field or an effect of asymmetry in the total electron content in the two hemispheres. We conclude that the Birkeland current magnitudes can be well described by combining estimates or measurements of the dayside reconnection rate and the ionospheric conductance.

**Acknowledgments**
J.C.C. was supported by a Science and Technology Funding Council (STFC) studentship (and, after this work was completed, a NERC grant NE/L007177/1). S.E.M. and J.A.C. were supported on STFC grant ST/K001000/1. The OMNI data were obtained from the GSFC/SPDF OMNIWeb interface at http://omniweb.gsfc.nasa.gov/, and the AMPERE data can be obtained from http://ampere.jhuapl.edu/. The authors acknowledge the International Space Science Institute, Bern, Switzerland, for hosting discussions useful to the study. J.C.C. would like to thank Simon Vaughan, Juan Hernandez Santisteban, and Christopher Boon for discussions on the statistical significance of higher correlations and Daniel Whiter for discussions on the hemispherical asymmetry in ionospheric conductivity.